# Investigation of the Cyprus donkey milk bacterial diversity by 16SrDNA high-throughput sequencing in a Cyprus donkey farm

P. Papademas[1],*[†], E. Kamilari[1][†], M. Aspri[1], D. A Anagnostopoulos[1], P. Mousikos[1], A. Kamilaris[2], and D. Tsaltas[1],*

[1] Department of Agricultural Sciences, Biotechnology and Food Science, Cyprus University of Technology, Limassol, Cyprus
[2] Faculty of Electrical Engineering, Mathematics and Computer Science (EEMCS) of the University of Twente, the Netherlands
* Correspondence: Photis Papademas: photis.papademas@cut.ac.cy, +357 2500 2581 (PP) & Dimitris Tsaltas: dimitris.tsaltas@cut.ac.cy, +357 2500 2545 (DT)
[†] P. Papademas and E. Kamilari contributed equally

**Interpretive Summary**
The interest in donkey milk is growing worldwide due to its functional and nutritional properties, especially for infants, immunocompromised and elderly. Therefore, the objective of this study was to identify and characterize the bacterial communities of donkey milk produced in a donkey farm in Cyprus by using culture-based approaches in combination with high-throughput sequencing. Results from this study confirms that the donkey milk bacterial microbiome is mostly comprised of Gram-negative bacteria. The findings from this study are expected to increase knowledge regarding the bacterial consortium of donkey milk and provide indications of the key bacterial microbiome that contributes to donkey milk's elevated nutritional value.

**Abstract:** The interest in milk originating from donkeys is growing worldwide due to its claimed functional and nutritional properties, especially for sensitive population groups, such as infants with cow's milk protein allergy. The current study aimed to assess the microbiological quality of donkey milk produced in a donkey farm in Cyprus using cultured-based and high-throughput sequencing (HTS) techniques. The culture-based microbiological analysis showed very low microbial counts, while important food-borne pathogens were not detected in any sample. In addition, HTS was applied to characterize the bacterial communities of donkey milk samples. Donkey milk was mostly comprised of: Gram-negative Proteobacteria, including *Sphingomonas*, *Pseudomonas Mesorhizobium* and *Acinetobacter*; lactic acid bacteria, including *Lactobacillus* and *Streptococcus*; the endospores forming *Clostridium*; and the environmental genera *Flavobacterium* and *Ralstonia*, detected in lower relative abundances. The results of the study support existing findings that donkey milk contains mostly Gram-negative bacteria. Moreover, it raises questions regarding the contribution: a) of antimicrobial agents (i.e. lysozyme, peptides) in shaping the microbial communities and b) of the bacterial microbiota to the functional value of donkey milk.

**Highlights**

- The metagenome of a Cyprus donkey milk was characterized by next generation sequencing.
- The Cyprus donkey milk was dominated by Gram-negative bacteria.
- Very low microbial counts and lack of food-borne pathogens were detected in donkey milk.

## 1. Introduction

Milk from non-traditional animal species (i.e., donkey, camel, and buffalo) are recently gaining interest for research and regulatory authorities, mainly because they are considered valuable alternative nutritional sources for specific population groups (i.e., infants, the elderly, immunocompromised, allergic to cow milk protein) (Jirillo, Jirillo and Magrone, 2010; Salimei and Fantuz, 2012; Aspri, Economou and Papademas, 2017). In particular, the interest in donkey milk has increased dramatically over the past few years due to its nutritional, nutraceutical, functional and immunological properties (Aspri, Economou and Papademas, 2017). Several studies have demonstrated that donkey milk maintains antimicrobial (Zhang *et al.*, 2008; Brumini *et al.*, 2013; Koutb, 2016; Adduci *et al.*, 2019), anti-inflammatory (Jirillo and Magrone, 2014; Yvon *et al.*, 2018), antimitotic, as well as antitumor (Mao *et al.*, 2009) capacities. Furthermore, it has been reported to be a suitable alternative for infants suffering from cow milk protein allergy (Souroullas, Aspri and Papademas, 2018).

Donkey milk is characterized by a very low microbial population which can be attributed to the increased concentrations of antimicrobial factors, including lysozyme and lactoferrin (Vincenzetti *et al.*, 2008; Tidona *et al.*, 2011). Lysozyme is an enzyme that catalyzes the hydrolysis of glycosidic bonds between N-acetylmuramic acid and N-acetyl-D-glucosamine residues of peptidoglycan, the primary component of the bacterial cell wall (Brumini *et al.*, 2016; Labella *et al.*, 2016). Lactoferrin is a multipurpose glycoprotein with bacteriostatic and bactericidal activities (Jahani, Shakiba and Jahani, 2015). Its antibacterial activity involves binding to lipopolysaccharide (LPS) of bacterial walls and: a) absorbs iron, which is required for bacterial growth (Ward and Conneely, 2004); b) prevents binding of important for bacterial pathogenesis compounds to LPS (Ochoa and Cleary, 2009); c) binds additional substances and compounds, including heparin, DNA, glycosaminoglycans, as well as metal ions such as $Mn^{3+}$, $Al^{3+}$, $Co^{3+}$, $Ga^{3+}$, $Zn^{2+}$, $Cu^{2+}$, etc. (Khan *et al.*, 2001); d) induces apoptosis in cells (Appelmelk *et al.*, 1994); and e) apolactoferrin (iron-free lactoferrin) damages the external membrane of Gram-negative bacteria by enhancing its permeability (Superti *et al.*, 2005). Indeed, the microbiological data of raw donkey milk shows a significantly low total bacteria count with a mean population of 2.40–5.87 log cfu/ml (Coppola *et al.*, 2002; Salimei *et al.*, 2004; Chiavari *et al.*, 2005; Zhang *et al.*, 2008; Malissiova *et al.*, 2016). However, the presence of food-borne pathogens such as *Escherichia coli* O157, *Listeria monocytogenes*, *Bacillus cereus*, and *Camplylobacter* spp., have been detected in some studies (Cavallarin *et al.*, 2015; Colavita *et al.*, 2016; Mottola *et al.*, 2018).

Despite the recognized benefits of donkey milk consumption, the existing microbial consortia and their possible contribution in the milk's nutritional value, have not been evaluated yet. Although various culture-dependent methodologies have identified the presence of bacteria, including food-borne pathogens, in aseptically collected milk, they don't suffice to provide complete information regarding several additional genera present in low numbers or difficult to be cultured (Quigley, O'Sullivan, *et al.*, 2013). Recently, the High Throughput Sequencing (HTS) technology, has been applied for deeper identification of the vastly diverse bacterial communities present in different types of milk (De Filippis, Parente and Ercolini, 2018; Oikonomou *et al.*, 2020). This technology provides the ability to characterize the microbiota present within a sample comprehensively, and it is characterized by increased sensitivity and high throughputness in comparison to other culture-independent methodologies. Amplicon sequencing achieves this by generating and sequencing in parallel thousands of specific DNA sequences, such as the bacterial 16S rDNA gene (Bokulich and Mills, 2013). The microbiome of donkey milk is hypothesized to be composed of bacteria commonly found in milk samples, but with adaptation to the elevated presence of antimicrobial compounds. Previous HTS studies on donkey milk bacterial communities identified increased relative representation of Gram-negative bacteria, such as *Pseudomonas* spp. (Soto del Rio *et al.*, 2017; Russo *et al.*, 2020).

Therefore, considering the growing interest in donkey's milk for infants, adults and elderly, the study aims to identify and characterize the bacterial communities of donkey milk produced in a donkey farm in Cyprus, as well as to evaluate its microbiological quality by using culture-based approaches in combination with Illumina MiSeq amplicon sequencing. The extracted findings are expected to increase knowledge regarding the bacterial consortium comprising the donkey milk and provide indications of the key bacterial microbiome that contributes to donkey milk's elevated nutritional value.

## 2. Materials and Methods

### 2.1. Collection of milk samples

Milk samples were collected from the "Golden Donkeys Farm", located in Larnaca District, Cyprus. All donkeys were fed the same diet consisting of hay, barley, corn, a concentrate of minerals, vitamins and salt following the European Directive 98/58/EC. Donkeys were healthy and no antibiotics were administrated prior to sampling. The process of milking was carried out in the stable and donkeys were milked manually from the same milker. Sampling for physicochemical and microbiological analysis was conducted weekly (33 weeks) from October of 2018 until May 2019 from the daily milk batch (20L from 20-25 milking donkeys). Milk samples for 16S rRNA gene amplicon analysis was conducted in May 2019 from eleven donkeys. For physicochemical and microbiological analysis from each donkey, a total of 250 ml of milk from both mammary glands were collected into a 250 ml sterile container. For 16S rRNA gene amplicon analysis a total of 100 ml of milk from both mammary glands were collected into two 50-ml sterile tubes (2 samples/donkey). During milking, the udder was cleaned using sterile wet wipes and the nipples using 70% ethanol and dried with sterile gauze. The donkeys were all multiparous. Donkey milk samples were placed in cool-boxes and immediately transported to the laboratory, where: a) they were kept at 4 °C and processed during the same working day for physicochemical and microbiological cultured-based analysis, or b) stored at −80 °C for 16S rRNA gene amplicon-HTS analysis.

### 2.2. Physicochemical analysis

Physicochemical analyses of fresh raw donkey milk were performed by using standard methods i.e. total nitrogen content (ISO 8968-1:2014), fat (ISO 488:2008) and total solids (ISO 6731:2010). All the analyses were done in triplicates, and average values were reported.

### 2.3. Microbiological Analysis

All samples were evaluated for total aerobic bacteria, Enterobacteriaceae, Staphylococci, *Escherichia coli and Bacillus cereus* after serial dilutions in saline solution (0.85% w/v), using pour or spread plate technique. Table 1 shows the growth media, incubation time, temperature and specific method used for each group of microorganisms inspected. *Listeria monocytogenes* analysis performed by using the method ISO 11290-1:2017. All the analyses were done in triplicates.

### 2.4. 16S rRNA gene DNA isolation

Five ml of donkey milk were mixed with forty-five ml of 2% tri-sodium citrate (Honeywell, Europe). After centrifugation at 16,000 × g for 5 min at 4°C, the top fat layer removed using sterile cotton swabs, and the supernatant discarded. Microbial DNA isolation was performed using DNeasy® PowerFood® Microbial Kit (MoBio Laboratories Inc., Carlsbad, CA, US), based on the manufacturer's instructions. The isolated DNA was kept at −20 °C until processing.

### 2.5. Quantification of total DNA

The total DNA extracted from the donkey milk samples was quantified fluorometrically using Qubit 4.0 fluorometer (Invitrogen, Carlsbad, CA) and Qubit dsDNA HS Assay Kit (Invitrogen). Evaluation of DNA purity achieved by measuring the ratios of absorbance A260/280 nm and A260/230 nm, using a spectrophotometer (NanoDrop Thermo Scientific, USA).

### 2.6. Barcoded Illumina MiSeq amplicon sequencing of bacterial 16s rRNA gene

A high-throughput sequencing approach was applied to isolated donkey milk DNA for the identification of the existing bacterial communities. The bacterial V3–V4 hyper-variable region of the 16S rDNA gene was amplified with the following 16S rDNA gene amplicon PCR primer pairs: a) forward primer (CCTACGGGNGGCWGCAG) and b) reverse primer (GACTACHVGGGTATCTAATCC), with the overhang adapter sequence addition. The paired-end approach based on the Illumina's protocol was applied (https://support.illumina.com/documents/documentation/chemistry_documentation/16s/16s-metagenomic-library-prep-guide-15044223-b.pdf) and as described by Kamilari et al. (Kamilari *et al.*, 2020). The quantification of each PCR product DNA concentration was performed using Qubit dsDNA High sensitivity assay. The estimation of DNA quality was evaluated using a bioanalyzer (Agilent 2200 TapeStation) (expected size ~550 bp). The purification of each PCR amplicon was performed using NucleoMag® NGS Clean-up and Size Select (Macherey-Nagel, Germany). Total amplicon products were normalized in equal concentrations and mixed in a single tube. The MiSeq 300 cycle Reagent Kit v2 (Illumina, USA) (5% PhiX) was applied for the sequencing runs, while the sequencing reaction was performed on a MiSeq Illumina sequencing platform.

*2.7. Bacterial Microbiome and Statistical analysis*

FASTQ sequences were analyzed using Qiime 2 version 2020.2 (Bolyen *et al.*, 2019). For quality filtering of raw reads the Phred33 quality threshold was applied. Adapter sequence removal, FASTQ trimming and read quality control performed using Trimmomatic (Bolger, Lohse and Usadel, 2014). Additionally, the DADA2 algorithm (Callahan *et al.*, 2016) performed correction of Illumina-sequenced amplicon errors, discarding reads with undesired quality and with more than 2 expected errors, as well as removing chimeric sequences. Sequences were aligned using Mafft (via q2-alignment) (Katoh and Standley, 2013). Alpha rarefaction analysis, alpha diversity metrics (Faith's phylogenetic diversity, Shannon, Inverse Simpson and Chao1) and beta diversity index (Bray Curtis similarity) were evaluated via the Qiime2 (version 2020.2) and primer e v7 (https://www.primer-e.com). Principle Coordinate Analysis (PCoA) were estimated using q2-diversity after 11 samples were rarefied (subsampled without replacement) to 77,143 sequences per sample. Alpha rarefaction curve was plotted with 25 sampling depths. The clustering of the 16S rDNA sequences and the filtering in Operational Taxonomic Unit (OTU) was performed using 16s Metagenomics App from BaseSpace against the Illumina-curated version of GreenGenes (v.05.2013) (DeSantis *et al.*, 2006; Klindworth *et al.*, 2013). The classified OTUs were defined at $\geq 97\%$ of sequence homology and converted to percentages (relative abundances), to determine the representation of each microbe among treatments. OTUs with relative abundance lower than 0.001% were excluded.

All raw sequence data in read-pairs format were deposited to the National Centre for Biotechnology Information (NCBI) in Sequence Read Archive (SRA) under BioProject PRJNA612663.

## 3. Results

*3.1. Physicochemical Analysis:*

The physico-chemical characteristics of fresh raw donkey milk were evaluated for the period October 2018-May 2019. Table 2 presents statistical values for each physicochemical parameter. Raw donkey milk was characterized by a mean protein content around 1.62 g/100ml and a mean fat content around 0.84 g/100ml. The mean dry matter observed in current donkey milk study was of 9.23 g/100ml.

*3.2. Culture-based microbiological analysis*

Table 3 presents the microbiological results of the 33 raw donkey milk samples for total viable microorganisms, Staphylococci, Enterobacteriacae, *Escherichia coli*, *Bacillus cereus* and *Listeria monocytogenes*. The mean value of viable microorganisms (TVC) was 3.80 log10 cfu/ml. Furthermore, Staphylococci and Enterobacteriacae were less than 4.7 log10 cfu/ml and 3.4 log10 cfu/ml respectively, while *Escherichia coli*, *Bacillus cereus* and *Listeria monocytogenes* were not detected.

*3.3. 16S rRNA gene amplicon-HTS analysis*

3.3.1. Abundance and diversity of members of the bacterial microbiota

Eleven (11) examined sample sets were used as input to the Illumina MiSeq to generate 281,294 high quality sequencing reads, with an average of 25,572.18 sequencing reads per sample (range = 17,413–35,159, STD = 5454.01) at the genus level (Table 4). High quality sequences were grouped into average number 357.91 OTUs (range = 266 - 492, SD =66.35). Shannon, Inverse Simpson, Chao1 and Chao2 estimators for genus level are also shown in Table 4.

Moreover, to evaluate differences in the bacterial community compositions of donkey milk samples, Weighted UniFrac distance-based microbiota structure analysis was performed. Bray-Curtis similarity index indicated increased similarity among the bacterial communities of milk samples in genus level (Table S1). PCoA of Bray-Curtis distance indicated no effective discrimination between samples (Figure 1). The principal coordinates 1, 2 and 3 explained 62.24%, 10.30% and 6.8% of the variation, respectively. The OTU network showed relation with changes in the microbial population and one main cluster was observed, from which the samples D6, D9 and D10 were discriminated.

3.3.2. Taxonomic composition of bacterial communities in Donkey milk samples

According to 16S rDNA sequencing, the bacterial communities of donkey milk consisted of mostly members of the phylum Proteobacteria. Members of the phyla Firmicutes, together with Bacteroidetes, Actinobacteria, Acidobacteria, Cyanobacteria and Verrucomicrobia were detected in lower relative abundances. Figure 2 illustrates the bacterial composition of the donkey milk samples based on the percentage of sequence reads identified at the genus levels. The most commonly detected bacteria, identified in percentages greater than 1% in all analyzed samples, were the Gram-negative bacteria *Sphingomonas* (16%-47%), *Pseudomonas* (8%-17%) and *Mesorhizobium* (11%-25%), as well as the genus *Acinetobacter,* which was detected in increased relative abundances in two samples (samples G3 and G6, 24% and 16% respectively). Moreover, Lactic acid bacteria (LAB), including the genera *Lactobacillus* and *Streptococcus,* were detected in relative abundances ranging from 1% to 4% in all samples tested. Additional commonly detected genera but in reduced relative abundances, included the genera *Ralstonia* (0.02% - 2.6%), *Aquabacter* (0 to 5%) and its phylogenetically related *Xanthobacter* (0% to 5.5%), as well as the proteolytic *Flavobacterium* (0 to 5%). Furthermore, number of reads representing 0.1% to 2% of the total reads per sample, of the spore-forming, butyrate-producing *Clostridium*, were also found.

## 4. Discussion

The current study is the first report in which HTS technology applied to investigate the bacterial communities of Cyprus donkey milk. 16S rRNA gene amplicon-HTS was used for an in-depth quantitative description of the bacterial population structure. Due to new information arising in recent years on the beneficial role of donkey milk consumption, such facilities are on a rise and milk production from other milk producing species is becoming a niche.

The results of the physicochemical parameters of donkey milk samples are in line with other studies (Guo *et al.*, 2007; Salimei and Fantuz, 2012; Malissiova *et al.*, 2016). The low content of donkey milk in fat is the main limitation for its use as the sole milk to children allergic to cow's milk protein during their first year of life since recommended dietary targets may not be achieved unless adequately supplemented with medium-chain triglycerides (D'Auria *et al.*, 2011; Salimei and Fantuz, 2012).

The microbiological quality of donkey milk using cultured based methods was in accordance with previous studies (Conte *et al.*, 2005, 2012; Pilla *et al.*, 2010; Sarno *et al.*, 2012; Cavallarin *et al.*, 2015; Malissiova *et al.*, 2016; Mottola *et al.*, 2018). In most studies, including this one, low bacteria counts have been observed. Moreover, only a few studies have shown the presence of some food-borne pathogens, but in the present study, no food-borne pathogens were detected. Furthermore, the low total population of viable microorganisms complies with the EC Regulation 853/2004, allowing the sale of donkey's milk under the clause "*other milk-producing species*," where the total bacterial plate count is less than 1,500,000 cfu/ml at 30 °C. Noteworthy, if raw milk from species other than cows is intended for the manufacture of products made with raw milk by a process that does not involve any heat treatment, food business operators must take steps to ensure that the raw milk used meets the following criterion: "Plate count at 30 °C (per ml) ≤ 500 000." (EC Regulation 853/2004). The high content of donkey milk in antimicrobial proteins, including lysozyme and lactoferrin, in combination with lactoperoxidase and immunoglobulins, are considered responsible for the low total bacterial counts (Salimei *et al.*, 2004; Vincenzetti *et al.*, 2008; Šarić *et al.*, 2012; Cosentino *et al,* 2016).

OTU analysis of the 16S rDNA gene sequences indicated that the Gram-negative bacteria *Sphingomonas*, *Mesorhizobium* and *Pseudomonas* were the most dominant genera detected in the Cyprus donkey milk samples. Other genera commonly occurred include *Acinetobacter*, *Lactobacillus*, *Streptococcus*, *Ralstonia*, *Clostridium* and *Flavobacterium*. Previous metagenomic studies in donkey milk microbiota have also detected the presence of these genera, except *Clostridium,* but in different relative abundances (Table 5) (Soto del Rio *et al.*, 2017; Russo *et al.*, 2020). Soto Del Rio et al. (Soto del Rio *et al.*, 2017) indicated that the predominant genera were *Pseudomonas, Ralstonia, Sphingobacterium, Acinetobacter, Cupriavidus, and Citrobacter,* although the core bacterial representation differed among samples. This is probably because the samples obtained from five different donkey dairy farms during two years, in contrast to the current study in which samples were obtained from one farm, during a shorter period. In agreement, Russo et al. (Russo *et al.*, 2020) identified also increased relative representation of the genus *Pseudomonas* in fresh donkey milk samples. Additional genera that detected in lower relative abundances included *Chryseobacterium*, *Sphingobacterium, Stenotrophomonas*, *Citrobacter* and *Delftia*. Similar 16S rDNA sequencing analyses in human and bovine milk also identified the frequent presence of the genera *Ralstonia, Sphingomonas* and *Pseudomonas* in all samples tested (Hunt *et al.*, 2011; Kuehn *et al.*, 2013).

The high abundance of Gram-negative compared to Gram-positive bacteria could be due to the presence of lysozyme. Donkey milk is characterized by a higher concentration of lysozyme (up to 4000mg/L) compared to bovine milk (0.09 mg/l) and human milk (up to 200 mg/L) (Chiavari *et al.*, 2005; Vincenzetti *et al.*, 2008). Its hydrolytic activity against the glycosidic bonds of peptidoglycan makes lysozyme more effective against Gram-positive bacteria. Indeed, recent 16S rDNA metagenomic studies that performed in other animals' milk with a lower concentration of lysozyme, including goat, sheep, cow and human, indicated the presence of Gram-positive bacteria such as *Staphylococcus* and *Streptococcus,* in high percentages *(*Table 6). Oikonomou et al., (Oikonomou *et al.*, 2020) reported that *Staphylococcus* and *Streptococcus* were among the most commonly detected genera between human and cow milk, in addition to *Corynebacterium, Pseudomonas*, *Bacteroides Bifidobacterium*, *Propionibacterium* and *Enterococcus.* Apart from lysozyme, a second antimicrobial agent that exists in donkey milk is lactoferrin. Lactoferrin is detected in lower concentration compared to lysozyme (up to 135 μg/ml) (Papademas *et al.*, 2019). These two proteins were reported to act synergistically against Gram-negative bacteria also (Ellison and Giehl, 1991; Hunt *et al.*, 2011). Lactoferrin's capacity to bind LPS of Gram-negative bacteria, may provide access to lysozyme's molecules to target and degrade the peptidoglycan in the cell wall. Based on the current findings, though, lactoferrin's presence in limited concentrations might not suffice to prevent the growth of Gram-negative bacteria.

The most abundant genus in all donkey milk samples was *Sphingomonas.* This genus is characterized by the presence of glycosphingolipids instead of lipopolysaccharides in their cell envelopes., Also, they possess the ability to grow in stressful for most bacteria environments (Nishiyama *et al.*, 1992; Krziwon *et al.*, 1995; Kelley *et al.*, 2004). They are considered spoilage bacteria in raw milk, in addition to *Acinetobacter* and spore-forming Clostridia (Zhang *et al.*, 2019). However, spoilage of milk has mostly been attributed to the psychrotrophic members of the genus *Pseudomonas. Pseudomonas* was detected in increased relative abundances in several milk samples (Table 7). The capability of *Pseudomonas* spp. to successfully utilize milk proteins and lipids due to their proteolytic and lipolytic enzymatic activities provides them with the ability to grow in raw milk (Quigley, McCarthy, *et al.*, 2013; Porcellato *et al.*, 2018). *Pseudomonas* spp. are considered to be among the species responsible for limiting donkey milk's self-life (approximately three days) (Soto del Rio *et al.*, 2017).

The genus *Mesorhizobium* comprises of soil bacteria that colonize legume roots and assist in the transformation of atmospheric nitrogen into plant-available compounds (Lindström *et al.*, 2010). Forage legumes are important sources of protein, fiber and energy for animal-based agriculture. Moreover, legumes grazing supports meat and milk production, as well as suppressed growth of internal parasites that provoke animals' mortality (Karaś *et al.*, 2015). Detection of *Mesorhizobium* was also reported in another donkey milk metagenomic study (Soto del Rio *et al.*, 2017) but in lower relative abundances. Additional nitrogen-fixing bacteria that detected, but in limited relative abundances (<1%), include *Rhizobium, Azorhizobium, Sinorhizobium/Ensifer*, *Azospirillum* and *Nitrospirillum*. The occurrence of these symbiotic bacteria might be associated with donkeys' nutrition, and specifically with legumes (Aganga, Letso and Aganga, 2000). Legumes are an essential source of donkeys' necessary amino acids since donkeys cannot store them efficiently. Rodriguez et al (Rodriguez *et al.*, 2014) in their work with human milk suggest that selected bacteria of the maternal GI microbiota can access the mammary glands through oromammary and enteromammary pathways. The mechanism involves dendritic cells and CD18 + cells, which would be able to take up nonpathogenic bacteria from the gut epithelia cells and subsequently, to carry them to other locations, including the lactating mammary gland.

The presence of LAB, including *Lactobacillus* and *Streptococcus,* was indicated in all donkey milk samples, with average relative abundance 2.42% (ranging from 1% to 4%). These results are in agreement with the other 16S rRNA gene amplicon HTS studies of (Soto del Rio *et al.*, 2017; Russo *et al.*, 2020) on donkey's milk microbiota, in which the average relative abundance of LAB were 4.2% and 2.55%, respectively. LAB are commonly detected in milk and dairy samples due to their capacity to ferment lactose successfully (Quigley, O'Sullivan, *et al.*, 2013; Kamilari *et al.*, 2019). Furthermore, the present of coccus-shaped bacteria such as *Streptococcus* may also be due to the high lysozyme content in donkey's milk. According to Neviani et al., (Neviani *et al.*, 1991) LAB cocci are more resistant to lysozyme than lactobacilli, and among lactobacilli, the lysozyme sensitivity is species or strain-specific; for instance, thermophilic species are more sensitive than hetero-fermentative mesophilic lactobacilli. LAB possess antimicrobial properties, mainly due to the presence of bacteriocin like inhibitory substances (Macaluso *et al.*, 2016). In a study, carried out by Aspri et al., (Aspri, Bozoudi, *et al.*, 2017; Aspri, Field, *et al.*, 2017) it was shown that the main LAB isolated from donkey milk samples belong to the *Enterococcus* species. Most of the isolates had interesting technological properties and were able to produce bacteriocins, while no pathogenicity was detected. Their presence in milk restricts the risk of food-borne diseases caused by bacteria, including *Staphylococcus aureus*, *Listeria monocytogenes*, pathogenic *Escherichia coli* and *Salmonella* spp. (Arqués *et al.*, 2015), and increases milk safety for consumers.

In the present study, the application of culture-based approaches identified the presence of Staphylococci and Enterobacteriacae in less than 4.7 log10 cfu/ml and 3.4 log10 cfu/ml, respectively. 16S rDNA sequencing analysis detected number of reads belonging to the genus *Staphylococcus* and the family Enterobacteriacae, that corresponded to low relative abundances (0.034% to 0.188% and 0.152% to 2.375 %, respectively). Moreover, the culture-based analysis indicated the absence of the species *Escherichia coli*, *Bacillus cereus* and *Listeria monocytogenes*. Regarding *Escherichia coli* and *Listeria monocytogenes,* culture-based results were in agreement with 16S rDNA metagenomic analysis findings. However, some reads corresponding to the genus *Escherichia/Shigella* were identified (5-52 reads) but assigned to the species *Escherichia/Shigella_dysenteriae* and *Escherichia_vulneris.* Also, HTS identified the presence of *Bacillus cereus* in two species, but in limited relative representation (0.02%). Based on these observations, 16S rRNA gene sequencing analysis showed higher sensitivity to detect specific bacteria than the usual culture-based method.

The present study indicated that donkey milk harbors complex bacterial communities containing different microorganisms. Still, the identification of the origin of the milk microbiota remains largely unknown. In agreement with other studies on donkey milk microbiome (Soto del Rio *et al.*, 2017; Russo *et al.*, 2020) the plethora of the detected bacteria are considered environmental. Their existence in donkey milk might come from external contamination of the breast (or udder) during nursing, derived from the mother's skin or the infant's oral cavity. This suggestion is supported by observations on human and bovine milk microbiota, in which species found in the teat skin (Doyle *et al.,* 2017), or the oral cavity (Cabrera-Rubio *et al*., 2012; Murphy *et al*. 2017; Williams *et al*. 2019) were detected in milk. In addition, some researchers suggest an endogenous origin of the milk microbiome, proposing the presence of a hypothetical enteromammary pathway (Jost *et al.,* 2015; Addis *et al.,* 2016, Williams *et al.,* 2019). Interestingly, in the present, as well as other studies on milk microbiota (see Table 5), the detection of LAB, such as *Lactobacillus* in milk, which contains species associated with the gut microbiome and not detected on the human and animals breast skin, supports this theory. Moreover, the presence of alive bacteria in the mammary gland of women who weren't breastfed before, indicates also that this tissue might not be sterile (Urbaniak *et al*., 2014). Noteworthy, bacteria associated with the animal's nutrition, such as *Mesorhizobium,* were also found in the present study; their origin though remains unspecified. Overall, more effort needs to be provided in order to specify whether the rich diversity of microbes detected in donkey milk is shaped due to the contamination of the environment or affected by the animal's nutrition and the microbial communities existed in the animal's gut.

## 5. Conclusions

This is the first study performed to characterize the bacterial communities of donkey milk in a Cyprus farm via HTS. It highlights and confirms that the donkey milk bacterial microbiome is mostly comprised of Gram-negative bacteria, possibly due to the increased concentration of lysozyme. In the future, additional donkey milk samples are to be analyzed to enable a broader evaluation and characterization of the existing bacterial communities. Factors that contribute to the conformation of donkey milk microbiota, such as the origin of milk, the environment, animals' health, diet, and genetics, will also be analyzed. The metagenomic analysis could be combined with other methodologies, including proteomics and metabolomics, for a sufficient estimation of the associations among the existing bacteria, the secreted metabolites and the antimicrobial agents detected in donkey milk. These analyses will shed more light on the nutritional benefit and antimicrobial activity of donkey milk.

**Author Contributions:** Conceptualization, D.T. and P.P.; methodology, E.K., D.A. and M.A.; formal analysis, E.K., M.A. and A.K.; investigation, E.K., D.A. and M.A.; resources, D.T., P.P. and P.M.; data curation, E.K.; writing—original draft preparation, E.K. and M.A.; writing—review and editing, E.K. D.T. and P.P.; supervision, D.T. and P.P.; project administration, D.T.; funding acquisition, D.T. All authors have read and agreed to the published version of the manuscript.

**Funding:** This research was funded by INTERREG Greece – Cyprus 2014-2020 Program (Project AGRO-ID, which is co-funded by the European Union (ERDF) and National Resources of Greece and Cyprus) and by the Research and Innovation Foundation (RIF) of Cyprus. Andreas Kamilaris has received funding from the European Union's Horizon 2020 Research and Innovation Programme under grant agreement No 739578 complemented by the Government of the Republic of Cyprus through the Directorate General for European Programmes, Coordination and Development

**Acknowledgments:** The authors would like to acknowledge the technical support and welcoming environment from Golden Donkeys Farm, Cyprus for the milk samples collection.

**Conflicts of Interest:** The authors declare no conflict of interest.

**Table 1.** Methods used for the enumeration of microorganisms.

| Group of microorganisms | Growth Media | Incubation Conditions | Reference Method |
|---|---|---|---|
| Total Aerobic Bacteria | PCA (Merck, Darmstadt, Germany) | 30 °C/ 72h | ISO 4833:2013 |
| Enterobacteriacae | VRBGA (Oxoid, Basingstoke, UK) | 37°C/24h | ISO 21528-2:2017 |
| Staphylococci | BP (Oxoid, Basingstoke, UK) | 37°C/48h | ISO 6888-1:1999 |
| *Bacillus cereus* | MYP (Merck, Darmstadt, Germany) | 30°C/48h | ISO 7932:2004 |
| *Escherichia coli* | TBX (Oxoid, Basingstoke, UK) | 44°C/24h | ISO 16649-2:2001 |
| *Listeria monocytogenes* | ALOA (Oxoid, Basingstoke, UK) | 37°C/48 h | ISO 11290-1:2017 |

**Table 2.** **Physico**chemical analysis of donkey milk samples (n=33).

| Chemical Parameter | Min | Max | Mean | SD |
|---|---|---|---|---|
| Fat (g/100ml) | 0.30 | 1.40 | 0.84 | 0.07 |
| Protein (g/100ml) | 1.30 | 1.96 | 1.62 | 0.05 |
| Total Solids (g/100ml) | 7.29 | 10.59 | 9.23 | 0.69 |

**Table 3.** Microbiological quality of donkey milk samples (n=33).

| Microbiological Parameters | Min | Max | Mean | SD |
|---|---|---|---|---|
| TVC (log cfu/ml) | 2.90 | 5.10 | 3.80 | 0.02 |
| Enterobacteriaceae (log cfu/ml) | <1.00 | 3.40 | 1.90 | 0.04 |
| Staphylococci (log cfu/ml) | <1.00 | 4.70 | 3.10 | 0.06 |
| *E. coli* (log cfu/ml) | <1.00 | <1.00 | <1.00 | <1.00 |
| *Bacillus cereus* (log cfu/ml) | <1.00 | <1.00 | <1.00 | <1.00 |
| *Listeria monocytogenes* | ND | ND | ND | ND |

**Table 4.** Sample information, microbial diversity and sequence abundance in genus level.

| Sample ID | Number of reads | Raw reads | Shannon | Simpson | Chao1 | Chao2 | Observed OTUs |
|---|---|---|---|---|---|---|---|
| D1 | 27133 | 37262 | 1.785 | 0.6832 | 472.61 | 263 | 266 |

| | | | | | | |
|---|---|---|---|---|---|---|
| D2 | 33039 | 45823 | 1.901 | 0.7308 | 636.81 | 632.96 | 328 |
| D3 | 24904 | 27634 | 2.097 | 0.7295 | 742.94 | 791.39 | 313 |
| D4 | 26371 | 37068 | 1.906 | 0.7289 | 873.48 | 1005.4 | 305 |
| D5 | 18898 | 28157 | 2.118 | 0.7521 | 938.9 | 1023.9 | 312 |
| D6 | 17413 | 24025 | 2.961 | 0.8594 | 1065.1 | 1218.2 | 358 |
| D7 | 24809 | 34292 | 2.218 | 0.7353 | 1177.9 | 1310.4 | 408 |
| D8 | 35159 | 49388 | 2.186 | 0.7795 | 1206 | 1290.8 | 410 |
| D9 | 27872 | 38799 | 3.368 | 0.9247 | 1236.5 | 1330.6 | 492 |
| D10 | 20160 | 28157 | 2.358 | 0.7534 | 1255 | 1393.2 | 418 |
| D11 | 25536 | 35190 | 2.199 | 0.7741 | 1273 | 1383.2 | 327 |

**Table 5.** The relative representation of bacterial genera that detected in milk samples via 16S rDNA sequencing.

| Type of milk | Country | Relative abundance | | | Reference |
|---|---|---|---|---|---|
| | | ≥25% | 10% - 24% | 1% - 9% | |
| Donkey (n=11) | Cyprus | *Sphingomonas* | *Mesorhizobium, Pseudomonas* | *Acinetobacter*, LAB (*Lactobacillus, Streptococcus*) *Ralstonia* | Present study |
| Donkey (n=11) | Italy | *Pseudomonas* | *Ralstonia* | *Acinetobacter, Citrobacter, Sphingobacterium, Cupriavidus, Stenotrophomonas*, LAB (*Carnobacterium, Enterococcus, Lactobacillus, Lactococcus, Leuconostoc, Streptococcus*) | (Soto del Rio *et al.*, 2017) |
| Donkey (n=10) | Italy | *Pseudomonas* | *Chryseobacterium* | *Stenotrophomonas, Sphingobacterium, Citrobacter, Delftia, Azospirillum, Massilia, Serratia* | (Russo *et al.*, 2020) |
| Sheep (n=37) | Spain | - | *Staphylococcus, Lactobacillus, Corynebacterium* | *Streptococcus, Escherichia/Shigella* | (Esteban-Blanco *et al.*, 2020) |
| Goat (n=10) | Cyprus | *Lactococcus, Leuconostoc* | *Pseudomonas* | *Carnobacterium Pahnella* | (Papademas *et al.*, 2019) |
| Goat (n=8) | United States | *Pseudomonas* | *Rhodococcus* | *Micrococcus, Stenotrophomonas, Phyllobacterium, Streptococcus, Agrobacterium* | (McInnis *et al.*, 2015) |
| Cow (n=27) | France | - | *Staphylococcus, Corynebacterium* | *Ruminococcus, Aerococcus, Bifidobacterium, Flacklamia, Jeotgalicoccus, Trichococcus* | (Falentin *et al.*, 2016) |
| Cow (n=48) | United States | - | *Staphylococcus* | *Streptococcus, Corynebacterium, Mycoplasma, Fusobacterium,* | (Lima *et al.*, 2017) |
| Cow (n=36) | United States | - | *Corynebacterium, Acinetobacter, Psychrobacter* | *Arthrobacter, Staphylococcus, Chryseobacterium, Coxiella, Facklamia, Prevotella, Pseudomonas,* | (Bonsaglia *et al.*, 2017) |

| Human (n=33) | Slovenia | *Staphylococcus* *Streptococcus* | - | *Treponema, Ruminobacter, Wautersiella, Cellvibrio, Ruminococcus, Aerococcus, Coprococcus, Clostridium, Bacteroides Acinetobacter, Gemella, Rothia Corynebacterium, Veillonella* | (Treven *et al.*, 2019) |
|---|---|---|---|---|---|
| Human (n=10) | Ireland | - | *Pseudomonas* | *Staphylococcus, Streptococcus, Elizabethkingia, Variovorax,* | (Murphy *et al.*, 2017) |
| Human (n=21) | Spain | *Staphylococcus* | *Pseudomonas* *Streptococcus* *Acinetobacter* | *Finegoldia, Corynebacterium* | (Boix-Amorós, Collado and Mira, 2016) |
| Human (n=21) | United States | *Streptococcus* | *Staphylococcus* | *Gemella, Veillonella, Rothia, Lactobacillus, Propionibacterium, Corynebacterium* | (Williams *et al.*, 2017) |
| Human (n=133) | China, Taiwan | - | *Streptococcus* *Pseudomonas* *Staphylococcus* | *Lactobacillus, Propionibacterium, Herbaspirillum, Rothia, Stenotrophomonas, Acinetobacter, Bacteroides, Halomonas* | (Li *et al.*, 2017) |
| Human (n=80) | China | *Staphylococcus* *Streptococcus* | *Pseudomonas* (in c-section) | *Ralstonia* | (Kumar *et al.*, 2016) |
| | South Africa | *Pseudomonas* | *Staphylococcus* | *Streptococcus* | |
| | Finland | *Staphylococcus* *Streptococcus* | - | *Ralstonia, Pseudomonas* (in c-section) | |
| | Spain (vaginal delivery) | *Staphylococcus* | *Streptococcus* | *Pseudomonas, Acinetobacter, Ralstonia,* | |
| | Spain (cesarean delivery) | *Pseudomonas* | *Streptococcus* | - | |

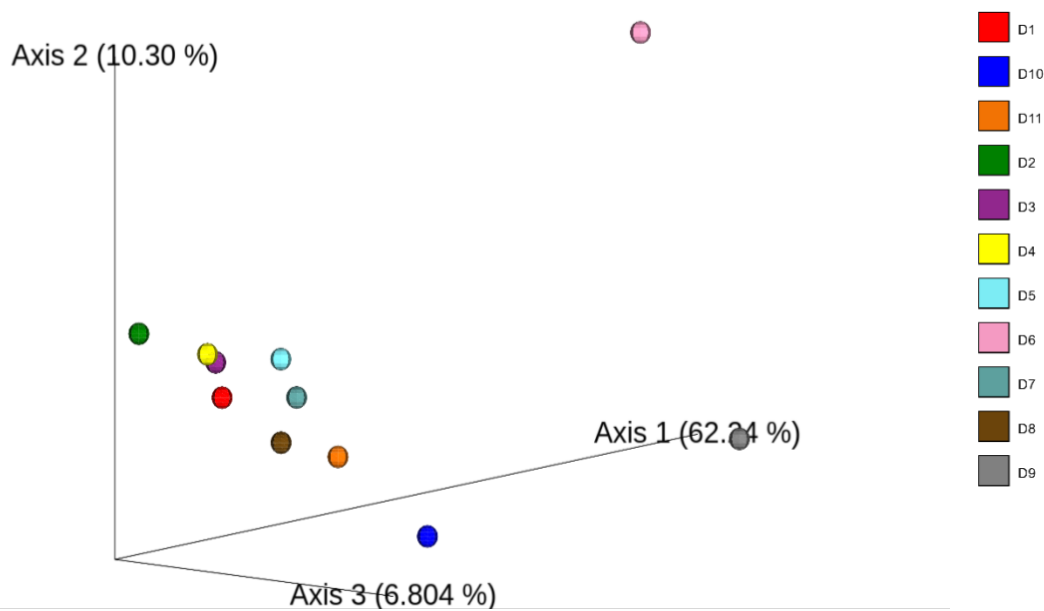

**Figure 1**. PCoA analysis of donkey milk samples. PCoA plots of Bray-Curtis distance. Clustering of points means similarity in relative abundances of OTUs among those samples.

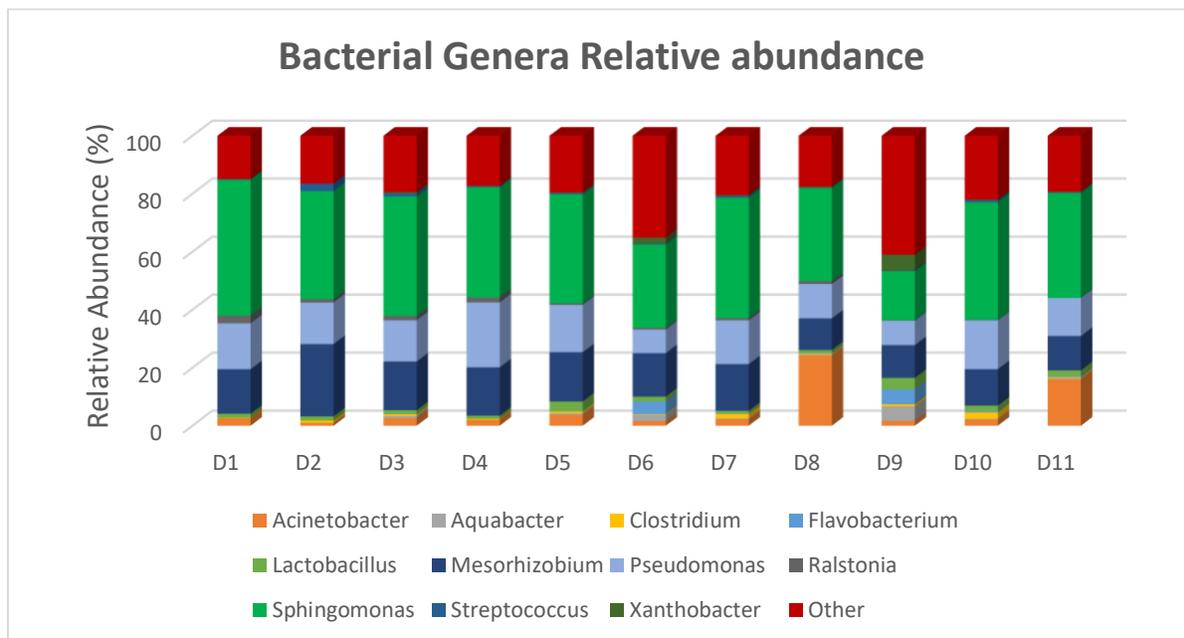

**Figure 2.** 3D 100% Stacked Column chart of the relative abundance of the major taxonomic groups detected by HTS at genus levels for 11 donkey milk samples. Only OTUs with an incidence above 1% in at least two samples are shown.

**Supplementary Materials:**

**Table S1**. Bray-Curtis distance matrix, donkey milk samples in genus level, all data combined.

|     | D1     | D2     | D3     | D4     | D5     | D6     | D7     | D8     | D9    | D10 |
|-----|--------|--------|--------|--------|--------|--------|--------|--------|-------|-----|
| D2  | 83.574 |        |        |        |        |        |        |        |       |     |
| D3  | 77.955 | 69.371 |        |        |        |        |        |        |       |     |
| D4  | 86.235 | 80.275 | 79.288 |        |        |        |        |        |       |     |
| D5  | 75.765 | 66.403 | 89.056 | 75.995 |        |        |        |        |       |     |
| D6  | 53.292 | 46.679 | 66.337 | 54.418 | 66.242 |        |        |        |       |     |
| D7  | 85.305 | 77.649 | 84.045 | 86.803 | 81.034 | 58.579 |        |        |       |     |
| D8  | 78.144 | 70.477 | 68.116 | 74.377 | 65.809 | 49.111 | 76.12  |        |       |     |
| D9  | 49.508 | 44.607 | 58.762 | 50.413 | 61.361 | 71.833 | 55.046 | 47.377 |       |     |
| D10 | 74.563 | 67.96  | 86.777 | 76.326 | 86.035 | 63.755 | 84.795 | 66.245 | 57.38 |     |

| D11 | 74.828 | 66.329 | 79.95 | 75.064 | 80.273 | 58.765 | 80.119 | 79.829 | 55.027 | 80.056 |